\documentclass[aps,prl,preprint,showpacs]{revtex4}

\usepackage{graphicx}

\newcommand{\us}{\mbox{$\mu$s}}
\newcommand{\trho}{\mbox{$\tilde{\rho}$}}
\newcommand{\Hn}{\mbox{$^{1}$H}}
\newcommand{\C}{\mbox{$^{12}$C}}

\newcommand{\N}{\mbox{$^{12}$N}}

\newcommand{\Bu}{\mbox{$^{11}$B}}
\newcommand{\numu}{\mbox{$\nu_{\mu}$}}
\newcommand{\numub}{\mbox{$\bar{\nu}_{\mu}$}}
\newcommand{\nue}{\mbox{$\nu_{e}$}}

\newcommand{\nueb}{\mbox{$\bar{\nu}_{e}$}}
\newcommand{\nutau}{\mbox{$\nu_{\tau}$}}
\newcommand{\nux}{\mbox{$\nu_{x}$}}
\newcommand{\ep}{\mbox{e$^{+}$}}

\newcommand{\el}{\mbox{e$^{-}$}}
\newcommand{\pos}{\mbox{e$^{+}$}}

\newcommand{\mup}{\mbox{$\mu^{+}$}}

\newcommand{\pip}{\mbox{$\pi^{+}$}}
\newcommand{\mupdecay}{\mbox{\mup\ $\rightarrow\:$ \pos $\!$ + \nue\ + \numub}}
\newcommand{\mupdeb}{\mbox{\mup\ $\rightarrow\:$ \pos $\!$ + \nueb\ + \numu}}
\newcommand{\mupnueb}{\mbox{\mup\ $\rightarrow\:$ \pos $\!$ + \nueb\ + $^{(}\bar{\nu}^{)}$}}
\newcommand{\mupeg}{\mbox{\mup\ $\rightarrow\:$ \pos $\!$ + $\gamma$}}
\newcommand{\pipmup}{\mbox{\pip $\rightarrow\:$ \mup + \numu}}
\newcommand{\CC}{\mbox{\C\,(\,\nue\,,\,\el\,)\,\N }}

\newcommand{\CBn}{\mbox{\C\,(\,\nueb\,,\,\ep\,n\,)\,\Bu }}
\newcommand{\CCprot}{\mbox{p\,(\,\nueb\,,\,\ep\,)\,n }}
\newcommand{\excl}{\mbox{\C\,(\,\nue\,,\,\el\,)\,\N$_{\rm g.s.}$}}
\newcommand{\nuex}{\mbox{\nue $\rightarrow\,$\nux }}

\newcommand{\numunue}{\mbox{\numu $\rightarrow\,$\nue }}
\newcommand{\numubnueb}{\mbox{\numub $\rightarrow\,$\nueb }}

\newcommand{\NCL}{\mbox{90\%\,C.L.}}
\newcommand{\Dm}{\mbox{$\Delta m^2$}}
\newcommand{\sit}{\mbox{$\sin ^2(2\Theta )$}}
\newcommand{\Gdng}{\mbox{Gd\,(\,n,$\gamma$\,)}}
\newcommand{\pn}{\mbox{p\,(\,n,$\gamma$\,)}}

\begin{document}

\title{Improved limits on \nueb\ emission from \mup\ decay}

\newcommand{\authoratik}{$^{a}$}
\newcommand{\authoratipe}{$^{b}$}
\newcommand{\authoratubonn}{$^{c}$}
\newcommand{\authoratuerl}{$^{d}$}
\newcommand{\authoratuka}{$^{e}$}
\newcommand{\authoratuoxf}{$^{f}$}
\newcommand{\authoratqmwc}{$^{g}$}

\newcommand{\addressofik}[1]{$^{a}$ #1 }
\newcommand{\addressofipe}[1]{$^{b}$ #1 }
\newcommand{\addressofubonn}[1]{$^{c}$ #1 }
\newcommand{\addressofuerl}[1]{$^{d}$ #1 }
\newcommand{\addressofuka}[1]{$^{e}$ #1 }
\newcommand{\addressofuoxf}[1]{$^{f}$ #1 }
\newcommand{\addressofqmwc}[1]{$^{g}$ #1 }

\author{
%
%
B.~Armbruster\authoratik, G.~Drexlin\authoratik,
K.~Eitel\authoratik, T.~Jannakos\authoratik,
J.~Kleinfeller\authoratik,
R.~Maschuw\authoratik$^{,}$\authoratubonn, C.~Oehler\authoratik,
P.~Plischke\authoratik, J.~Reichenbacher\authoratik,
M.~Steidl\authoratik, B.~Zeitnitz\authoratik$^{,}$\authoratuka,
H.~Gemmeke\authoratipe, M.~Kleifges\authoratipe,
C.~Eichner\authoratubonn, C.~Ruf\authoratubonn,
B.A.~Bodmann\authoratuerl,
E.~Finckh\authoratuerl, J.~H\"o\ss
l\authoratuerl,
P.~J\"unger\authoratuerl,
W.~Kretschmer\authoratuerl,
J.~Wolf\authoratuka,
N.E.~Booth\authoratuoxf,
I.M.~Blair\authoratqmwc, J.A.~Edgington\authoratqmwc \\
\smallskip
\smallskip
\footnotesize \it
\addressofik{Institut f\"ur Kernphysik,
  Forschungszentrum Karlsruhe, 76021 Karlsruhe, Germany}\\
\addressofipe{Institut f\"ur Prozessdatenverarbeitung und Elektronik,
  Forschungszentrum Karlsruhe, 76021 Karlsruhe, Germany}\\
\addressofubonn{Institut f\"ur Strahlen- und Kernphysik, Universit\"at Bonn,
  Nu\ss allee 14-16, 53115 Bonn, Germany}\\
\addressofuerl{Physikalisches Institut, Universit\"at Erlangen-N\"urnberg,
  Erwin-Rommel-Str.~1, 91058 Erlangen, Germany}\\
\addressofuka{Institut f\"ur experimentelle Kernphysik, Universit\"at Karlsruhe,
  Gaede-Str.~1, 76128 Karlsruhe, Germany}\\
\addressofuoxf{Department of Physics, University of Oxford,
  Keble Road, Oxford OX1 3RH, United Kingdom}\\
  \addressofqmwc{Physics Department, Queen Mary, University London,
  Mile End Road, London E1 4NS, United Kingdom}\\
} \affiliation{ }

\date{\today}

\begin{abstract}
We investigated \mup\ decays at rest produced at the ISIS beam
stop target. Lepton flavor (LF) conservation has been tested by
searching for \nueb\ via the detection reaction \CCprot . No
\nueb\ signal from LF violating \mup\ decays was identified. We
extract upper limits of the branching ratio for the LF violating
decay \mupnueb\ compared to the Standard Model (SM) \mupdecay\
decay: $BR < 0.9(1.7)\cdot 10^{-3}$ (\NCL) depending on the
spectral distribution of \nueb\ characterized by the Michel
parameter $\trho=0.75 (0.0)$ . These results improve earlier
limits by one order of magnitude and restrict extensions of the SM
in which \nueb\ emission from \mup\ decay is allowed with
considerable strength. The decay \mupdeb\ as source for the \nueb\
signal observed in the LSND experiment can be excluded.
\end{abstract}


\maketitle

\section{Introduction \label{intro}}

In the Standard Model (SM), the main decay mode of positive muons
is the decay into a positron and two neutrinos $\mu^+ \rightarrow
e^+ + \nu + \nu'$. Assuming conservation of the additive lepton
family or flavor (LF) numbers $L_e$ and $L_{\mu}$, the neutrino
flavors are fixed to be $\nu=\nue$ and $\nu'=\numub$. The
neutrinos are massless with the \nue\ being a left-handed
neutrino, the \numub\ a right-handed anti-neutrino. The structure
of the muon decay can be described by the V--A theory of weak
interactions. Therefore $\mu$ decay as a purely leptonic process
has been used to study with high precision the SM of weak
interactions. The Lorentzian V--A structure of the \mup\ decay can
be tested by measuring the massive leptons, i.e. the initial \mup\
and the final \pos\ \cite{fetscher} or by investigating the
neutrino energy spectrum \cite{karmen_om}. However, to test
conservation of the LF numbers $L_e$ and $L_{\mu}$ in $\mu$ decay
it is essential to observe the final neutrino states
\cite{langacker}. All tests so far show no deviations from the SM.

However, the LF number violating decay mode \mupdeb\ is allowed in
many extensions of the SM, e.g. left--right (LR) symmetric models
\cite{moh75,her92,her92b,moh93}, GUT models with dileptonic gauge
bosons \cite{fujii}, extensions involving additional scalar
multiplets \cite{Pakv02} or supersymmetric models with R parity
violation \cite{halprin}, together with the LF number violating
decay \mupeg\ \cite{carlos}. Although the energy scale of LR
symmetry of weak interactions or the appearance of supersymmetric
particles is expected to be in the range of 0.1--1\,TeV, precision
measurements at intermediate energies can provide essential
restraints on the parameters used in various models. Therefore,
the detailed investigation of the \mup\ decay plays a major role
in determining the structure of weak interactions and the
precision of lepton number conservation.

On the other hand, there are clear evidences for neutrino
oscillations from experiments on atmospheric, solar and reactor
neutrinos \cite{Fogli02}. Since $\nu$ oscillations violate the
conservation of the lepton family numbers, such results enhance
the interest of searching for direct LF number violation. In
addition, there is a positive \nueb\ signal from the accelerator
experiment LSND \cite{LSNDfinal} which could be explained a priori
as an indication for \numubnueb\ oscillations of \numub\ from
\mupdecay\ or directly for the decay mode \mupnueb . Due to
limited statistics and energy resolution, this ambiguity is not
resolved by the LSND experiment itself.

The spallation source ISIS at the Rutherford Laboratory, UK, is a
unique source of \mup\ to study such decays. The {\bf KA}rlsruhe
{\bf R}utherford {\bf M}edium {\bf E}nergy {\bf N}eutrino
experiment investigated the neutrinos produced at ISIS through the
decays of \pip\ and \mup\ at rest. One purpose of the KARMEN
experiment was the investigation of $\nu$--nucleus interactions on
\C\ \cite{karmen_nc2}. The good agreement of the measurements with
theoretical predictions allowed a sensitive search for processes
forbidden in the SM such as $\nu$--oscillations, \numunue\ and
\numubnueb\ in the appearance mode \cite{karmen_final} and \nuex\
in the disappearance mode \cite{karmen_dis} or non--SM decay modes
of \pip\ and \mup .

In this letter we report the results of the search for \nueb\ from
\mup\ decay at rest (DAR). Note that \nueb\ from non SM
interactions can be produced by either \mupnueb\ in the ISIS
target or by oscillations \numubnueb\ of \numub\ on their way to
the detector with \numub\ being produced at ISIS in SM \mup\
decays. While the \nueb\ energy spectrum is fixed in the DAR
\mupnueb\ with a spatial flux according to a $r^{-2}$ dependence,
the energy and spatial distributions of \nueb\ from oscillations
strongly depend on the oscillation parameters, i.e. the mass
difference $\Delta m^2_{ij} = |m_i^2-m_j^2|$. With its excellent
energy resolution of $\sigma_E/E=11\%/\sqrt{E[{\rm MeV}]}$, the
KARMEN detector is able to separate different scenarios for
potential \nueb\ occurrence. Although the search for oscillations
\numubnueb\ and for the decay \mupnueb\ use the same detection
reaction \CCprot\ for \nueb , the different physics and
consequently the different \pos\ spectral distributions result in
two separate analyses.

\section{\protect\nueb\ from LF violating \mup\ decays \label{motivation}}

In the SM, applying the V--A theory, the energy spectra of
massless neutrinos from \mup\ decay \mupdecay\ can be calculated
neglecting radiative corrections as \cite{mich}
  \begin{equation}
        \rm{N}(\epsilon)d\epsilon \propto \epsilon^{2}\,[\,3(1-\epsilon)\,+\,
        \frac{2}{3}\rho(4\epsilon-3)\,]d\epsilon
  \label{michel}
  \end{equation}
with the relative energy
$\epsilon=\rm{E}_{\nu}/\rm{E}_{\rm{max}}$,
$\rm{E}_{\rm{max}}=52.83$\,MeV for the decay at rest, and the
Michel parameter $\rho = 0(0.75)$ for \nue(\numub), respectively.

Looking for physics beyond the SM, \nueb 's from \mupnueb\ in
general have non-zero mass and contributions of left- or
right-handed chirality eigenstates. As only the \nueb\ in the
decay \mupnueb\ is identified in the experiment, the second
emitted (anti)\-neutrino $^{(}\bar{\nu}^{)}$ is not determined.
Since our experimental result sets upper limits on \mupnueb ,
these limits also apply for the specific case
$^{(}\bar{\nu}^{)}=\numu$, which is the dominant one for certain
model assumptions \cite{her98}.

Taking the actual direct mass limits for \nueb ,
$m(\nueb)<2.2$\,eV \cite{Mainz} (and hence for \numu\ and \nutau\
masses through the mixing manifested in the experiments on
neutrino oscillations), the $\nu$ masses are very small compared
to the mass of the charged leptons or the energy scale of the
neutrinos from \mup\ decays at rest. Assuming Majorana type
neutrinos, the \nue\ of left-handed chirality emitted in the SM
decay \mupdecay\ could be detected via the reaction \CCprot\ since
there is no distinction between \nue\ and \nueb\ . However, the
detection of \nueb 's emitted in muon decays with left-handed
helicity would be strongly helicity--suppressed ($1-\beta\approx
o(10^{-14})$ for a neutrino of 10\,MeV energy and a rest mass of
2\,eV/c$^2$) since only right-handed anti-neutrinos are absorbed
via \CCprot . In the case of Dirac type neutrinos, the above
argument applies for left-handed chirality states of the \nueb\
emitted. Therefore, KARMEN as any detector up to date is sensitive
only to right-handed \nueb 's.

With the rest masses much smaller than the energy of all neutrinos
emitted in \mupnueb , an analytical description of the neutrino
spectra similar to the one in equ.~(\ref{michel}) can be applied,
with the spectral parameter \trho\ to be specified, replacing the
SM Michel parameter $\rho$. In some SM extensions with \mupdeb ,
the \nueb\ and \numu\ take the places of the SM \numub\ and \nue ,
respectively, with $\trho(\nueb)=0.75$ \cite{her92}. In others,
$\trho=0$ for the emitted \nueb\ \cite{Pakv02}. In our analysis,
we therefore investigate the \nueb\ emission for a variety of
\trho\ parameters.

\section{Experimental Configuration and Data evaluation \label{experiment}}

The experiment was performed at the neutrino source of the ISIS
synchrotron which accelerates protons to an energy of 800\,MeV
before striking a massive beam stop target. On average, $4.59\cdot
10^{-2}$ \pip\ per incident proton are produced which are stopped
within the target and decay at rest. Neutrinos emerge
isotropically from the consecutive decays at rest (DAR) \pipmup\
and \mupdecay\ \cite{bob} assuming the $\nu$--flavors of the SM
decay channels. Neutrinos from \mup\ DAR have a continuous energy
spectrum according to equ.~(\ref{michel}). Due to the narrow time
structure of 525\,ns of the proton pulses muons are produced in a
short time window compared to their lifetime of 2.2\,\us.

The neutrinos are detected in a 56\,t scintillation calorimeter
\cite{karmen_det} at a mean distance of 17.6\,m from the ISIS
target. The calorimeter is a mineral oil based scintillator
segmented into 512 independent modules. Gadolinium within the
module walls allows effective neutron detection via \Gdng\ in
addition to the capture on the hydrogen of the scintillator via
\pn . The scintillation detector provides an almost pure target of
\C\ and \Hn\ for $\nu$-interactions. Three veto layers ensure a
search for LF violating \mup\ decays almost free of cosmic
background.

\nueb 's from \mup\ decay can be detected via the $(\pos,n)$
sequence from charged current reactions \CCprot\ and \CBn . Hence,
the signature is a prompt \pos\ and a delayed, spatially
correlated $\gamma$ signal from the capture of the thermalized
neutron by \pn\ with $E_{\gamma}=2.2$\,MeV or \Gdng\ with
$\sum\,E_{\gamma}=8$\,MeV. The flux averaged (taking
equ.~(\ref{michel}) with $\rho=0.75$) cross section of \CCprot\ is
$\sigma = 93.5\cdot 10^{-42}$\,cm$^2$ \cite{beacom}. The \CBn\
contribution to $(\pos,n)$ sequences has a cross section of
$\sigma = 8.52\cdot 10^{-42}$\,cm$^2$ \cite{kolbe} which is
further reduced relative to \CCprot\ by the abundance ratio
H/C=1.767 within the scintillator.

\begin{figure}
 \includegraphics[width=8.5cm]{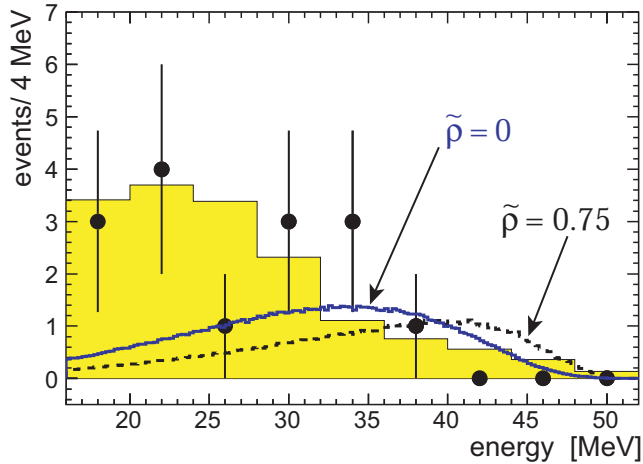}
 \caption{Visible energy distribution of candidate events
with background expectation (shaded area). The solid and dashed
lines show the \NCL\ limit for an additional \nueb\ signal with a
spectral parameter $\trho=0$ and $\trho=0.75$,
respectively.\label{pos_spectra}}
 \end{figure}

A positron candidate is accepted only if there is no activity in
the central detector or in the veto system up to 24\,\us\
beforehand. The prompt event is searched for in an interval of
0.6\,\us\ to 10.6\,\us\ after beam-on-target. The time structure
of the prompt \pos\ event relative to the proton pulses has to
follow the \mup\ decay time constant of 2.2\,\us . The expected
visible \pos\ energy has been simulated in detail based on \nueb\
spectra with different values of the parameter \trho\ including
both detection reactions \CCprot\ and \CBn . As a result, the
prompt energy is required to be within 16\,MeV $\le$ E(prompt)
$\le$ 50\,MeV (see Fig.~\ref{pos_spectra}).

The time difference between the \pos\ and the capture $\gamma$ is
given by the thermalization and capture of neutrons and can be
approximated by an exponential with a time constant of
$\tau_n\approx 120$\,\us . Therefore, the delayed event has to
appear within 1.3\,m$^3$ around the prompt event position,
correlated in time (5\,\us\ $\le \Delta t \le 300$\,\us) with a
visible energy E(delayed)$\le 8$\,MeV. For further details of the
data reduction of sequential event signatures and of the neutron
detection in KARMEN see also \cite{karmen_final}.

The raw data investigated in this search were recorded in the
measuring period of February~1997 to March 2001 and represent the
entire KARMEN\,2 data set which corresponds to 9425\,C protons on
target with $2.7\cdot 10^{21}$ \mup\ decays in the ISIS target.
Applying all evaluation cuts, 15 candidate sequences remain with
prompt energies as shown in Figure~\ref{pos_spectra}. The expected
background amounts to $15.8\pm 0.5$ events. This number comprises
3.9$\pm$0.2 events from cosmic induced sequences as well as $\nu$
induced reactions such as intrinsic source contamination of \nueb\
(2.0$\pm$0.2), $\nue$ induced random coincidences (4.8$\pm$0.3)
and (\el ,\ep) sequences from \excl\ with subsequent \N\ decay
(5.1$\pm$0.2). Except for the intrinsic \nueb\ contamination,
which has been deduced from detailed MC simulations, all the
background components have been measured in different time and
energy regimes with the KARMEN detector and extrapolated into the
evaluation cuts applied for this \nueb\ search.

\section{Results and Discussion \label{analysis}}

The expected number of \nueb\ induced events from \mupnueb\ is
determined by the detection efficiencies of the prompt positron
and the delayed neutron. The overall detection efficiency for
positrons is given in Table~\ref{tab_rhos} for a set of different
spectral parameters \trho\ including the contribution from \CBn ,
which effectively amounts to less than 5\% of the \CCprot\ signal
in the energy interval of 16--50\,MeV.
 \begin{table*}
 \caption{Flux averaged cross section $\langle \sigma \rangle$ for \CCprot\ and \CBn ,
 total efficiency for \pos\ detection, expected $(\pos,n)$ sequences for \mup\
 decaying entirely via \mupnueb , experimental results for potential
 \nueb-induced events and deduced upper limits for the branching ratio
 for different spectral parameters \trho . \label{tab_rhos}}
 \begin{ruledtabular}
 \begin{tabular}{l|c|c|c|c|c|c|c}
 \trho & \multicolumn{2}{c|}{$\langle \sigma \rangle [10^{-42}{\rm cm}^2]$} &
 \pos\ efficiency & N(\nueb)$_{\rm BR=1}$ &
 N(\nueb)$_{\rm best fit}$ & N(\nueb)$_{\rm 90\%CL}$ & BR (\NCL) \\
 & $\nueb\ + p$ & $\nueb\ + \C$ &[16-50]\,MeV &&&& \\ \hline
0.0 & $72.0$&$4.5$ & 0.450 & $4304\pm 403$ & $+0.3$ & $<7.1$ & $<1.7 \cdot 10^{-3}$ \\
0.25 & $78.8$&$5.8$ & 0.452 & $4773\pm 445$ & $-0.1$ & $<6.2$ & $<1.3 \cdot 10^{-3}$ \\
0.5 & $86.0$&$7.2$ & 0.456 & $5273\pm 489$ & $-0.4$ & $<6.0$ & $<1.1 \cdot 10^{-3}$ \\
0.75 & $93.5$&$8.5$ & 0.462 & $5828\pm 538$ & $-0.8$ & $<5.3$ & $<0.9 \cdot 10^{-3}$ \\
 \end{tabular}
 \end{ruledtabular}
 \end{table*}

Based on the Poisson statistics of the numbers of candidate events
and expected background, one can extract an energy-independent
upper limit for an additional signal \cite{pdg2002} of
N(\nueb)$<7.4$ excess events. However, there is additional
spectral information, as can be seen from
Figure~\ref{pos_spectra}. To use this, we applied a maximum
likelihood analysis varying the strength of a \nueb\ signal with
the energy distribution according to a set of different \trho\
parameters. Table~\ref{tab_rhos} shows the signal strength
N(\nueb)$_{\rm best fit}$ from the likelihood method. To extract
\NCL\ intervals for N(\nueb), we performed large samples of MC
simulations reproducing experiment-like spectra under different
signal hypotheses. Our experimental result is consistent with no
\nueb\ emission from \mup\ decay with upper limits given in
Table~\ref{tab_rhos}, extracted within a unified frequentist
analysis near the physical boundary N(\nueb)=0 following
\cite{feldcous}.

With a potential signal strength of N(\nueb)$=5828\pm 538$ for a
branching ratio $BR=1$ for LF number violating decays and
$\trho(\nueb)=0.75$, we set an upper limit of the branching ratio
of
  \begin{eqnarray} \nonumber
        BR &=& \frac{\Gamma(\mupnueb; \trho(\nueb)=0.75)}{\Gamma(\mupdecay)}
        < 9 \cdot 10^{-4} \nonumber
  \label{equ_BR}
  \end{eqnarray}
with \NCL\ as well as the upper limits given in
Table~\ref{tab_rhos} for other values of $\trho(\nueb)$.
Figure~\ref{pos_spectra} shows the visible \pos\ energies from
\mupnueb\ for two different \trho\ parameters with total strength
excluded at \NCL .

The above limits on the branching ratio $BR$ on \mup\ decays
emitting \nueb\ improve by more than an order of magnitude the
most sensitive limit so far of $BR(\mupdeb) < 0.012$ obtained by
the E645 experiment at LAMPF \cite{pdg2002,E645}.

In models extending the SM, the LF violating muon decay \mupdeb\
is often related to other LF violating processes, e.g. $\mu
\rightarrow 3e$, $\tau \rightarrow \mu e e$ or
muonium--antimuonium ($M\overline{M}$) conversion. Therefore,
limits such as the limit on the probability for spontaneous
conversion $P(M\rightarrow \overline{M})<8.2\cdot 10^{-11}$ (\NCL)
\cite{will} set also stringent limits on the coupling constants
responsible for the decay \mupdeb\ \cite{her98,Berg}.

The most conservative upper limit of $BR<1.7\cdot 10^{-3}$ for any
parameter \trho\ derived here is in direct experimental
disagreement with the possibility that the beam excess of \nueb\
seen in the LSND experiment with a branching ratio or probability
of $P=(2.64\pm 0.67\pm 0.45)\cdot 10^{-3}$ \cite{LSNDfinal} is due
to \mup\ decays with \nueb\ emission. In the LSND maximum
likelihood analysis of the data, the best fit to the data is found
to be an oscillation contribution with the parameters
$\Dm=1.2$\,eV$^2$ and $\sit=0.003$ \cite{LSNDfinal}. In this
analysis, oscillation events arise from \CCprot\ with \nueb\ via
\numubnueb\ from \mup\ DAR as well as from \CC\ with \nue\ via
\numunue\ from \pip\ decays in flight (DIF). For large \Dm , about
30\,\% of the oscillation signal is due to \numunue\ from DIF,
leading to a complex superposition of energy distributions for the
prompt events distorted by two different $L/E$ combinations, with
$L$ being the distance source--detector and $E$ the neutrino
energy.

To be able to compare our result quantitatively with the LSND
evidence, we follow the detailed statistical analysis of the LSND
data described in \cite{Joint}. In this analysis, a special cut
had been applied to select \nueb\ via \numubnueb\ from \mup\ DAR
only with almost no oscillation events from \numunue\ from \pip\
DIF. For large differences of the squared mass eigenvalues
$\Dm=100$\,eV$^2$, the extracted interval of the mixing amplitude
from the parameter region of \NCL\ was $3.8\cdot
10^{-3}<\sit<8.0\cdot 10^{-3}$, with the best fit at
$\sit=5.8\cdot 10^{-3}$ corresponding to 65.8(1.3) \nueb(\nue)
from \numubnueb(\numunue), respectively. For such large values of
\Dm , the energy spectrum of \nueb\ is in good approximation the
one given in equ.~(\ref{michel}) with $\trho=0.75$ \footnote{For a
quantitative comparison of the results for other values of \trho ,
one would need a specific re-analysis of the LSND data.}. Assuming
100\,\% decay probability with a spectral parameter $\trho=0.75$,
one would expect 22692 \nueb\ events seen in the LSND detector.
Taking the $65.8$ \nueb\ events deduced by the maximum likelihood
analysis as emitted in the decay \mupnueb\ , we extract a
branching ratio of $BR_{\rm LSND}(\trho=0.75)=2.9\cdot 10^{-3}$
with an approximated \NCL\ interval of $1.9\cdot 10^{-3}<BR_{\rm
LSND}<4.0\cdot 10^{-3}$. Our derived \NCL\ limit of $BR<0.9\cdot
10^{-3}$ clearly excludes the LSND interval of same confidence and
underlines the incompatibility of the two experimental results
interpretated in terms of LF violating \mup\ decays.

\begin{acknowledgments}
We acknowledge financial support from the German Bundesministerium
f\"ur Bildung und Forschung, the Particle Physics and Astronomy
Research Council and the Council for the Central Laboratory of the
Research Councils. We thank the Rutherford Appleton Laboratory and
the ISIS neutron facility for their hospitality.
\end{acknowledgments}

\end{document}